\documentclass{ws-procs9x6}

\def\beq{\begin{equation}}
\def\eeq{\end{equation}}

\def\epm#1#2{\hbox{${\lower1pt\hbox{$\scriptstyle +#1$}}
\atop {\raise1pt\hbox{$\scriptstyle -#2$}}$}}

\def\gsim{\mathrel{\rlap{\lower4pt\hbox{\hskip1pt$\sim$}}
    \raise1pt\hbox{$>$}}}         

\def\frac#1#2{{{#1}\over {#2}}}
\def\half{\hbox{${1\over 2}$}}

\def\as{\alpha_s}

\catcode`@=11 
\def\slash#1{\mathord{\mathpalette\c@ncel#1}}
 \def\c@ncel#1#2{\ooalign{$\hfil#1\mkern1mu/\hfil$\crcr$#1#2$}}
\def\lsim{\mathrel{\mathpalette\@versim<}}
\def\gsim{\mathrel{\mathpalette\@versim>}}
 \def\@versim#1#2{\lower0.2ex\vbox{\baselineskip\z@skip\lineskip\z@skip
       \lineskiplimit\z@\ialign{$\m@th#1\hfil##$\crcr#2\crcr\sim\crcr}}}
\catcode`@=12 

\def\be{\begin{equation}}
\def\ee{\end{equation}}
\def\bea{\begin{eqnarray}}
\def\eea{\end{eqnarray}}

\def\epm#1#2{\hbox{${\lower1pt\hbox{$\scriptstyle +#1$}}
\atop {\raise1pt\hbox{$\scriptstyle -#2$}}$}}
\def\wup{{W^+}}

\begin{document}

\title{\hfill {\rm IFUM-818/FT}\medskip\\
Polarized Structure Functions\\ with Neutrino Beams}

\author{Stefano FORTE}

\address{ Dipartimento di Fisica, Universit\`a di Milano and\\ 
INFN, Sezione di Milano, Via Celoria 16, I-20133 Milano, Italy}
\maketitle

\abstracts{We review the potential impact of neutrino data on the
  determination of the spin structure of the nucleon. We show that a
  flavour decomposition of the parton structure of the nucleon  as required by
  present-day precision phenomenology 
   could only be achieved at a neutrino factory. 
We discuss how neutrino scattering data would allow a
  full resolution of the nucleon spin problem.}
\begin{center}
 Invited plenary talk at\\
SPIN 2004\\
Trieste, November 2004\\
{\it to be published in the proceedings}
\end{center}
\bigskip
\bigskip\bigskip
\section{Physics with neutrino beams}
Physics with neutrino beams has played a crucial role in establishing
the standard model and its structure, specifically in leading to the discovery of
weak neutral currents and of their
properties.\cite{alvaro} Currently, while the physics {\it of}
neutrinos gives  us the first evidence of physics
beyond the standard model\cite{guido}, the use of neutrinos as probes
appears to be the only way of accessing subtle details of the
structure of the standard model, on the one hand, and of the nucleon,
on the other hand. This is due to the obvious fact that
weak currents, unlike the electromagnetic current, couple nontrivially
to spin and flavour. 

Current data offer tantalizing evidence
of this situation: neutrino scattering data
from the NuTeV collaboration\cite{nutev} 
provide evidence for unexpected effects, either
in the standard model, or in the structure of the
nucleon.\cite{dfgrs}
However, existing beams are insufficient to
exploit the potential of neutrino probes, because of scarce
intensity and lack of control of the beam spectrum, due to the fact
that neutrinos are obtained from the decay of secondary beams (pions). 
This has recently led to several proposals of facilities where
neutrinos would be produced as decay products of a primary beam:
either muons (neutrino factories~\cite{nufact}) 
or radioactive nuclei ($\beta$--beams~\cite{bbeam} ). This would allow
the production of $\sim10^{20}$ neutrinos/year (neutrino factory) or
$\sim10^{18}$ neutrinos/year ($\beta$-beam) with full control of the
energy spectrum, to be compared to $\sim10^{16}$ neutrinos/year of
present-day experiments. The prospects for the construction of such
facilities, which are being studied in Europe, Japan and the U.S.A, 
have recently improved, in particular due to a renewed commitment of CERN towards
future neutrino facilities.\cite{villars} 

\section{Deep-inelastic scattering with neutrino beams}
\subsection{Structure functions and parton distributions}
Inclusive DIS is the standard way of accessing
the parton content of hadrons. The use of neutrino beams allows one to
study DIS mediated by the weak, rather than electromagnetic interaction.
 The neutrino-nucleon deep-inelastic
cross section for charged--current interactions, up to
corrections suppressed by powers of $m_p^2/Q^2$ is given by
\bea
\label{disxsec}
&&
\frac{d^2\sigma^{\lambda_p\lambda_\ell}(x,y,Q^2)}{dx dy}
=
\frac{G^2_F}{  2\pi (1+Q^2/m_W^2)^2}
\frac{Q^2}{ xy}\Bigg\{
\left[-\lambda_\ell\, y \left(1-\frac{y}{2}\right) x { F_3(x,Q^2)}
      \right.\nonumber\\ &&\quad\left.
+(1-y) { F_2(x,Q^2)} + y^2 x {
F_1(x,Q^2)}\right]
 -2\lambda_p
  \left[
     -\lambda_\ell\, y (2-y)  x { g_1(x,Q^2)}
\right.\nonumber\\ &&\qquad\left. -(1-y) {
g_4(x,Q^2)}- y^2 x { g_5(x,Q^2)}
  \right]
\Bigg\},
\eea
where $\lambda$ are the lepton and proton helicities (assuming
longitudinal proton polarization), and the
kinematic variables are $y=\frac{p\cdot q}{p\cdot
k}$ (lepton fractional energy loss), $x= \frac{Q^2}{2 p\cdot
q}$ (Bjorken $x$). 
The neutral--current cross--section is found from Eq.~(\ref{disxsec})
by letting
$m_W\to m_Z$ and multiplying by an
overall factor  $[\half(g_V-\lambda_\ell
g_A)]^2$.

The advantage of $W$ and $Z$-mediated DIS over conventional 
$\gamma^*$ DIS  is clear when inspecting the
parton content of the polarized and unpolarized structure functions
$F_i$ and $g_i$. Up to $O(\alpha_s)$ corrections, in terms of the
unpolarized and polarized quark distribution for the $i$--th flavor
 $q_i\equiv
q_i^{\uparrow\uparrow}+q_i^{\uparrow\downarrow}$ and 
$\Delta q_i\equiv 
q_i^{\uparrow\uparrow}-q_i^{\uparrow\downarrow}$
\begin{center}
\begin{tabular}[c]{ccc}

NC&$F_1^{\gamma} =\half\sum_{i}  e^2_i\left(q_i+\bar
q_i\right)$\quad\qquad&$ g_1^{\gamma}=\half\sum_{i}
e^2_i\left(\Delta q_i+\Delta \bar
q_i\right)$\\
NC&$F_1^{Z} =\half\sum_{i}  (g^2_V+g^2_A)_i\left(q_i+\bar
q_i\right)$\quad\qquad&$ g_1^{Z}=\half\sum_{i} (g^2_V+g^2_A)_i
\left(\Delta q_i+\Delta \bar
q_i\right)$\\
NC&$F_3^{Z} =2\sum_{i}  (g_Vg_A)_i \left(q_i+\bar
q_i\right)$\quad\qquad&$ g_5^{Z}=-\sum_{i}(g_Vg_A)_i
\left(\Delta q_i+\Delta \bar
q_i\right)$\\
CC&{ $F_1^\wup =\bar u + d + s + \bar c$}\quad\qquad&${ g_1^\wup=\Delta\bar u + \Delta d +
\Delta s + \Delta \bar c}$\\
CC&${ -F_3^\wup/2 = \bar u - d - s +\bar c }$\quad\qquad & { $g_5^\wup = \Delta \bar u -\Delta d -\Delta s +\Delta\bar c$}\\
\phantom{CC}& $F_2= 2 x F_1$& $g_4= 2 x g_5$\\
\end{tabular}
\end{center}
\vspace*{-8pt}
Here $e_i$ are the electric charges and $(g_V)_i$, $(g_A)_i$ are the weak
charges of the $i$--th quark flavor.
If $W^+\to W^-$ (incoming $\bar \nu$ beam), then
$u\leftrightarrow d,\>c\leftrightarrow s$. 
Of course, beyond leading order in the strong coupling each
quark or antiquark flavor's contribution receives $O(\as)$
corrections proportional to itself and to all other quark, antiquark
and gluon distributions. However, the gluon correction is flavor--blind, and
thus decouples from the parity--violating structure functions $F_3$,
$g_4$ and $g_5$. 

\begin{figure}[ht]
\includegraphics[width=.49\linewidth,clip]{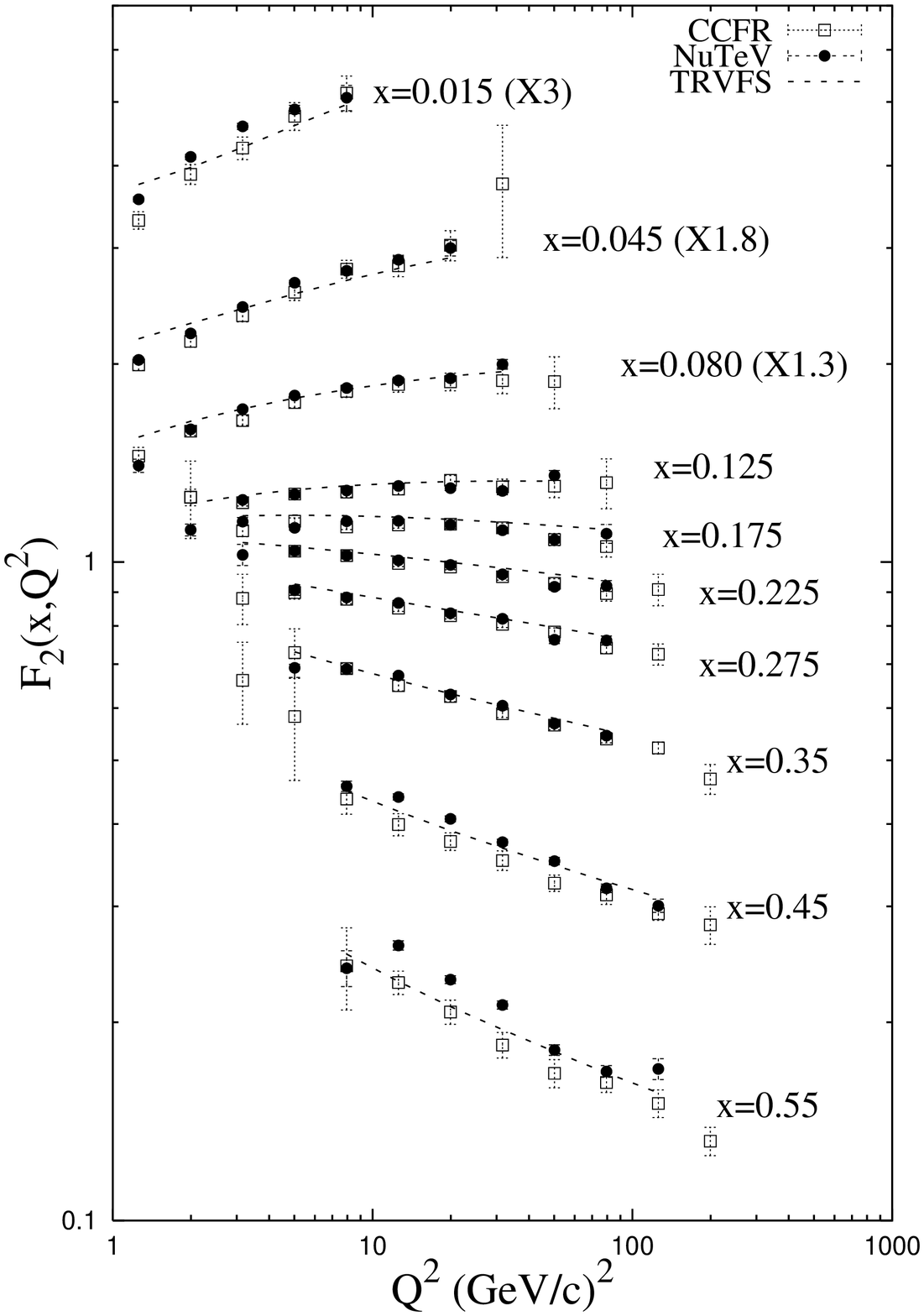}
\includegraphics[width=.49\linewidth,clip]{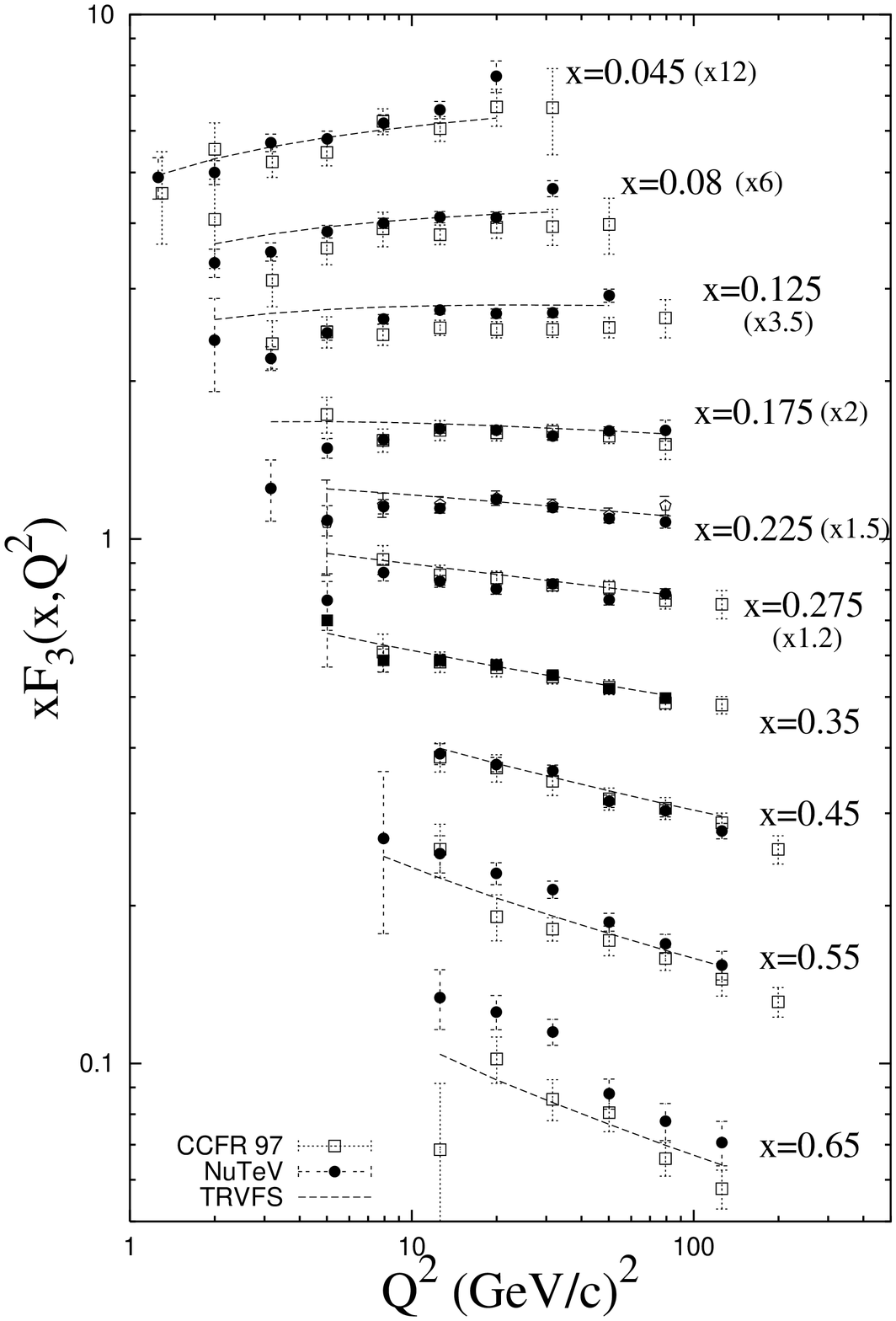}
\vskip-.25cm
\caption{Current unpolarized proton structure functions from $\mu$
  scattering (from ref.~[10]). The structure functions
  shown are $\nu+\bar\nu$ averages.\label{strfun}}
\vspace*{-12pt}
\end{figure}
In neutral current DIS only the
C-even combinations
$q_i+\bar q_i$ and $\Delta q_i+\Delta\bar q_i$ are accessible.
Furthermore, 
the structure functions
$F_3$, $g_4$ and $g_5$ are parity--violating, and therefore not
accessible in virtual photon scattering. 
In the presence of weak couplings, more independent linear
combination of individual quark and antiquark distributions are
accessible, thereby allowing one to disentangle individual flavours
and antiflavours.~\cite{bhm,fmr} 
\subsection{Current data and future prospects}

Structure function measurements with neutrino beams have been
performed recently by the CCFR/NuTeV collaboration,\cite{nutevdis}
while DIS results have been announced, but not yet published, by the
CHORUS\cite{chorus} and NOMAD\cite{nomad} collaborations. Older
results, including historic bubble--chamber data, have been reanalized
and collected in ref.\cite{BPZ}. The recent NuTeV
structure function data, based on a sample of $8.6\times 10^5$ $\nu$
and $2.3\times 10^5$ $\bar \nu$ DIS events, have led to reasonably
precise determinations of the structure functions $F_2$ and $F_3$ 
(see fig.\ref{strfun}). However, they still have rather lower
accuracy than charged-lepton DIS data. Furthermore, 
only $\nu+\bar\nu$ structure function averages are  determined:
 $F_3^{\nu}+F_3^{\bar\nu}$ (compare eq.~(\ref{disxsec})),
 from the difference
$\sigma^\nu-\sigma^{\bar \nu}$, and
$F_2^{\nu}+F_2^{\bar\nu}$  from the 
 average $\sigma^\nu+\sigma^{\bar \nu}$ using a
   theoretical determination of $F_3$. Hence, results are not free from
   theoretical assumptions, and only some linear combinations of
   parton distributions are
   accessible. Finally, present-day neutrino target--detectors are very large
   in order to ensure reasonable rates with available beams: e.g. the
   NuTeV target-calorimeter consists of 84 iron
   plates, 10~m$\times$10~m$\times$10~cm. Clearly, this prevents the
   possibility of putting the target inside a polarizing magnet.

\begin{figure}[ht]
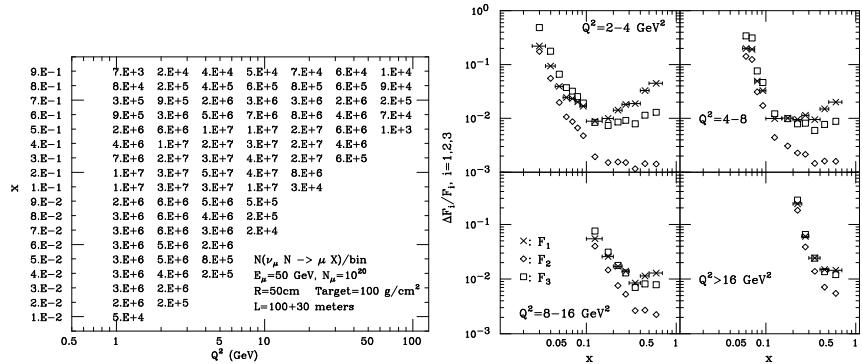

\includegraphics[width=.49\linewidth,clip]{xqrates.eps}
\includegraphics[width=.49\linewidth,clip]{F123err.eps}
\vskip-.25cm
\caption{Unpolarized DIS event rates (left) and errors on unpolarized
  parton distributions (right) for one year of running 
in the CERN neutrino factory scenario
  (from ref.~[14]
). \label{nufunp}}
\vspace*{-4pt}
\end{figure}

At a neutrino factory (in the CERN scenario\cite{nufact,cernufact})
the neutrino beam originates from the decay
of $10^{20}$  $\mu$ per year, stored in a 50~GeV ring. With
a target
effective density of $100$~g/cm$^2$, corresponding to, say, a 10~m
long deuterium  target\cite{bhm} and a radius of 50~cm, one could then
count on
$5\times10^8$ DIS events per year, with the rates in individual bins
shown in fig.\ref{nufunp}. Furthermore, taking
advantage of the fact that
$y=Q^2/(2 x m_p E_\nu)$, at fixed $x$ and $Q^2$, $y$ varies with
the incoming $\nu$ energy.
Because the beam at a neutrino factory is broad-band, if 
the kinematics of the DIS event can be fully reconstructed on an
event-by-event basis, it is
then possible to disentangle the contributions of the
individual structure functions to the cross section Eq.~(\ref{strfun}) by fitting
the $y$ dependence of the data for fixed $x$ and $Q^2$. The errors on  individual
structure functions obtained through such a procedure  are shown in
fig.\ref{nufunp}, and are in fact rather smaller than those on current
charged-lepton DIS structure functions.

Recent results of the NuTeV collaboration on the CC/NC total neutrino
DIS cross sections highlight the potential and limitations of current
neutrino data. The NuTeV data\cite{nutev} allow a determination of
the Paschos-Wolfenstein\cite{PW} ratio
\bea
R^-&=&\frac{\sigma_{NC}({\nu})-\sigma_{NC}(\bar{\nu})}{
\sigma_{CC}({\nu})-\sigma_{CC}(\bar{\nu})}\nonumber\\
&=&\left(\frac12 -\sin^2\theta_W\right)
+ 2
\left[{\frac{(u-\bar u)- (d-\bar d)}{u-\bar u+ d-\bar
    d}} -\frac{s-\bar s}{u-\bar u+ d-\bar d}
\right]\times 
\\ &&\Bigg[ \left(\frac12 -
 \frac76\sin^2\theta_W\right)
+{4\over 9} {\alpha_s\over 2\pi} \left(\frac12 -\sin^2\theta_W\right) +O(\as^2)\Bigg]
+O(\delta (u-d)^2,\>\delta s^2)\nonumber
\label{pweq}
 \eea
where $u$, $d$, $s$, etc. denote the second moments of the
corresponding parton distributions. All dependence on parton
distributions disappears assuming isospin for an isoscalar  target,
if one also assumes $s=\bar s$ (which is nontrivial for second
moments).

Using these assumptions, NuTeV has arrived at a
determination of $\sin^2\theta_W$ which differs by about three sigma
from the current standard best-fit. This indicates either physics
beyond the standard model, or a nucleon
structure which is subtler than expected.\cite{dfgrs} 
Small violations of
isospin are produced by QED corrections, and in fact phenomenological evidence
supports an isospin violation which reduces the observed discrepancy
by about one $\sigma$.\cite{iso} Also, a global fit
to world data  favors a strangeness asymmetry of the size and
magnitude required to remove the effect entirely.\cite{strange} 
However, in both cases the null assumption which leads to the
discrepancy with the standard model cannot be really excluded 
within the required precision.
Whereas less inclusive data, such as $W$ production at
hadron colliders,  might give us some extra handle on the
flavor decomposition of parton distributions,\cite{unpolrev} only a
neutrino factory would allow a full determination of the flavour and
antiflavour content of the nucleon\cite{cernufact} as required for
this kind of precision physics.

\section{Polarized physics with neutrino beams}
\subsection{Polarized DIS and the proton spin puzzle}

The determination of the polarized structure of the nucleon has
progressed considerabily since the surprizing discovery of the
smallness of the nucleon's singlet axial charge
$a_0$.\cite{polrev} The focus of current phenomenological
activity has shifted from inclusive deep-inelastic scattering to less
inclusive data, largely driven by the desire to access
quantities, such as transversity, which decouple from inclusive
DIS. Whereas semi-inclusive data give us some hint of the flavour
structure of the nucleon,\cite{seminc} experience in the
unpolarized case\cite{unpolrev} teaches that they cannot compete in
accuracy with DIS data, while hadron collider data play a
complementary role.

 \begin{figure}[ht]
\begin{minipage}[t]{.49\linewidth}
\includegraphics[width=.9\linewidth,clip]{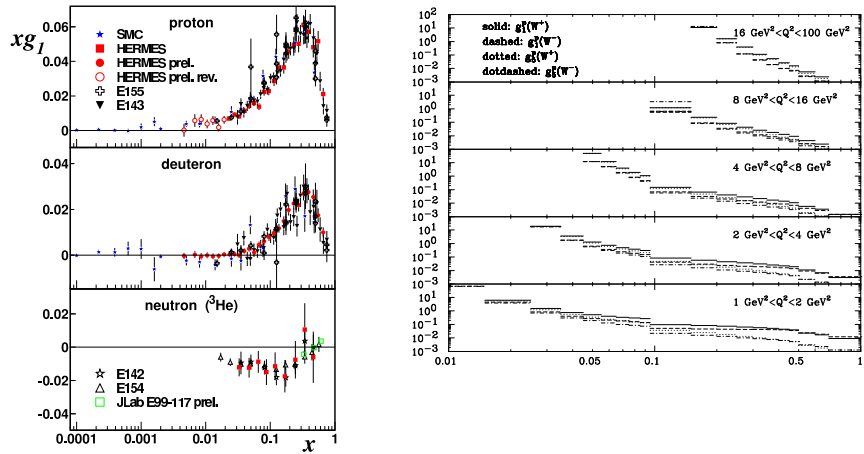}
\end{minipage}
\begin{minipage}[t]{.49\linewidth}
\includegraphics[width=1.15\linewidth,clip]{perr.ps} 
\end{minipage}
\vskip-.8cm
\caption{Current polarized structure function data (left, from
  ref.~[19]
) and expected accuracies in the CERN neutrino
  factory scenario (right, from ref.~[14]
).\label{nufpol}}
\end{figure}

However, a detailed understanding of the flavour and antiflavour
content of the nucleon is mandatory if one wishes to elucidate its spin
structure.\cite{polthrev} Indeed, the reason why the smallness of the singlet axial
charge $a_0$ is surprizing is that it signals a departure from the
quark model, in which  $a_0$ is the total quark spin fraction, and 
$a_0\approx a_8$, the difference being
due
to the strange contribution, which is expected to be small by the
Zweig rule. The octet axial charge $a_8$ cannot in practice be determined from
neutral-current DIS data, hence
it is currently determined using SU(3) 
from baryon $\beta$--decay constants: $a_8=0.6\pm 30\%$, while
$a_0=0.10\epm{0.17}{0.11}$ (at $Q^2=\infty$).
Whereas $a_8$ does not
depend on $Q^2$, $a_0$ does:
because of the
axial anomaly, $\partial_\mu j^\mu_5\not=0$.
It turns out, however, that it is possible to choose a
factorization scheme in such a way that the quark distribution is
scale--independent. In such case $a_0$ and the total quark spin
fraction $\Delta \Sigma$ are
no longer equal:
\bea\label{anomeq}
a_3&=&  \Delta u+\Delta \bar u-(\Delta d+\Delta \bar d)\nonumber\\
a_8&=& \Delta u+\Delta \bar u+\Delta d+\Delta \bar d - 2  (\Delta
s+\Delta\bar s) \nonumber\\
\Delta \Sigma&=& \Delta u+\Delta \bar u+\Delta d+\Delta \bar d+\Delta s+\Delta \bar s\\
a_0&=& \Delta \Sigma-\frac{n_f\as}{2\pi} \Delta G\nonumber
\eea
 where $\Delta q_i$ and $\Delta G$ are respectively the quark and
gluon spin fractions.

One can then envisage various scenarios for the nucleon spin content.\cite{polthrev,fmr}
A first possibility is that 
perhaps,  even though $a_0$ is small, $\Delta
\Sigma$ eq.~(\ref{anomeq}) is large, because $\alpha \Delta g$ is large
(`anomaly' scenario). If instead $\alpha \Delta g$ is small, there are
two possibilities. 
Either the determination of $a_8$ from octet decays using
isospin is incorrect, the Zweig rule in actual fact holds,
$a_0\approx a_8$, 
and the strange spin fraction $\Delta s+\Delta\bar s$ is 
small. Else, $\Delta s+\Delta\bar s$ is large, i.e. comparable
to $\Delta u+\Delta\bar u$ and $\Delta d+\Delta\bar d$. This is
predicted for instance to happen in instanton models or in Skyrme
models. These two cases, however, predict respectively that $\Delta s$
and $\Delta \bar s$ are separately large (`instanton' scenario) or
that $\Delta s\ll\Delta \bar s$ (`skyrmion' scenario). In short
different models of the nucleon spin structure lead to
distinct predictions for its polarized content.

\begin{center}\begin{table}[ph]
\tbl{Quark and gluon first moments. Both statistical and systematic
  errors are given for current values (from ref.~[23])
  statistical only for the neutrino factory scenarios (from ref.~[9]).}
{\footnotesize
\begin{tabular}{@{}|c|r||rrr|@{}}
\hline
{} &{} &{} &{} &{}\\[-1.5ex]
{} & present &  anomaly & instanton &  skyrmion\\[1ex]
\hline
{} &{} &{} &{} &{}\\[-1.5ex]
$\Delta g$ &$0.8\pm0.2\pm 0.4$
&$0.86\pm 0.10$   & $0.20\pm 0.06$  & $0.24\pm 0.08$
\\[1ex]
$\Delta \Sigma$ & $0.38\pm0.03\pm0.04$
&$0.39\pm 0.01$   &$0.321\pm 0.006$ & $0.324\pm 0.008$\\[1ex]
$a_3$ &$1.11\pm0.04\pm0.04$ 
&$1.097\pm 0.006$ &$1.052\pm 0.013$ & $1.066\pm 0.014$\\[1ex]
$a_8$ &$0.6\pm0.2(?)$         &$0.557\pm 0.011$ &$0.572\pm 0.013$ & $0.580\pm 0.012$\\[1ex]
$\Delta s-\Delta\bar s$ & ?
&$-0.075\pm 0.008$&$-0.007\pm 0.007$&$-0.106\pm 0.008$\\[1ex] 
\hline
\end{tabular}\label{table1} }
\vspace*{-13pt}
\end{table}\end{center}
Current data provide a fairly accurate determination of
$g_1^\gamma(x,Q^2)$ (figure ~3). From these only the 
C-even combination $\Delta
q_i+\Delta \bar q_i$ can be determined, while $\Delta g$ can be
extracted from scaling violations, albeit with large
errors.\cite{abfr} A reasonably accurate determination of the
isotriplet component is then possible, especially its first moment. 
However, the individual valence flavours can be
determined much less accurately, partly because of their admixture with
the gluon eq.~(\ref{anomeq}). In fact, only first moments can be
determined with reasonable accuracy (table 1), while  the shape of 
individual parton distributions is only known rather poorly (figure~4).

\subsection{The spin of the nucleon at a neutrino factory}
At a neutrino factory, significant rates could be achieved with small
targets:\cite{bhm} even with a rather conservative effective density
of $10$~g/cm$^2$,
with a detector radius of 50~cm,
the structure functions $g_1$,  $g_5$ could still be
independently measured with an accuracy which is about one order of
magnitude better than that with which $g_1$ is determined 
in present charged lepton DIS experiments (figure~3).
On the basis of such data, the flavour structure of the nucleon could
be entirely disentangled:\cite{abfr} in particular, the difference of any two
flavour or antiflavour fractions could be determined, typically
with uncertainties of order of 1\% (table~1).\\

\begin{figure}[ht]
\includegraphics[width=.49\linewidth,clip]{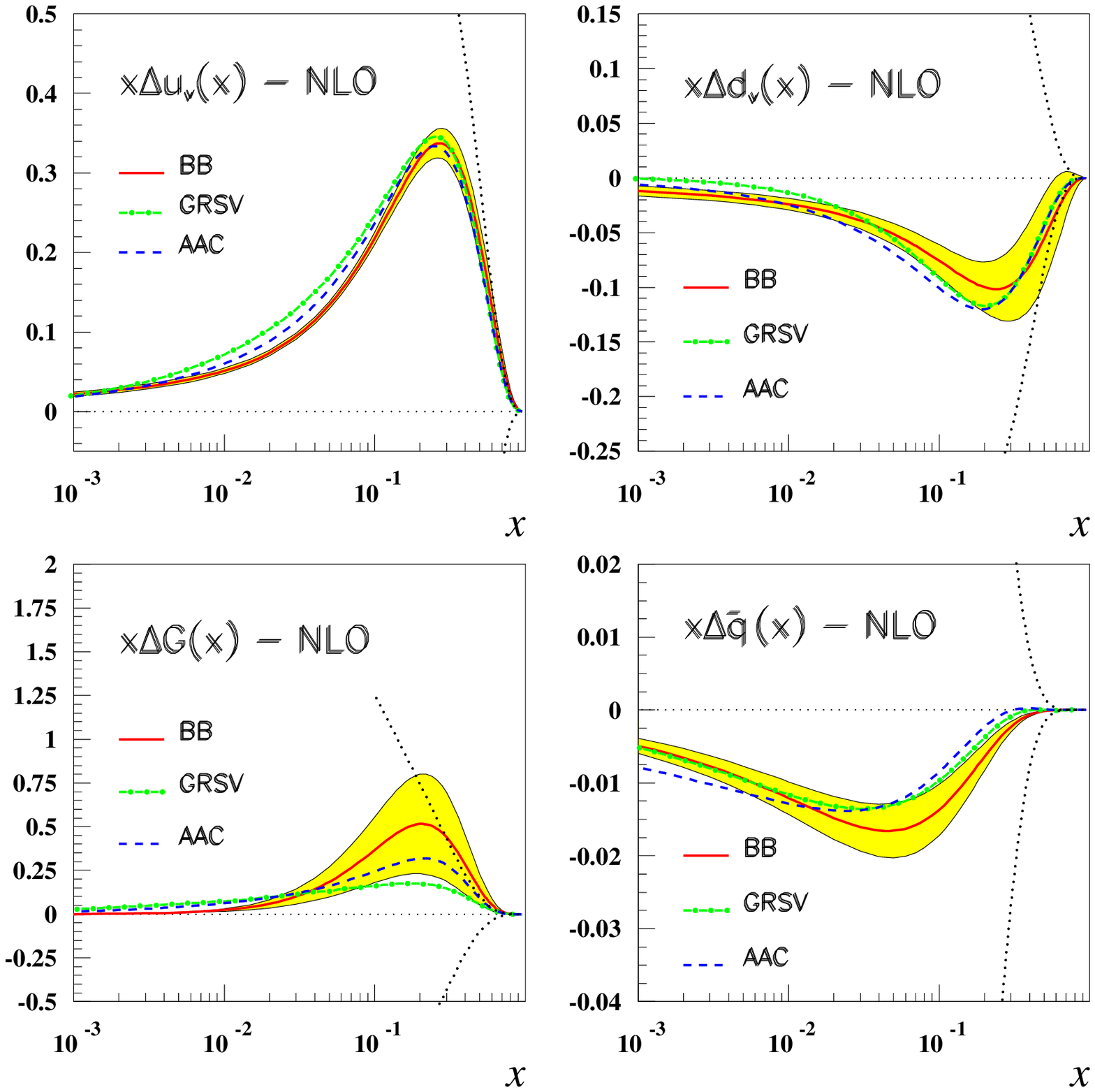}
\begin{minipage}[b]{.49\linewidth}
\includegraphics[width=.7\linewidth,clip]{deltau.ps}\\
\includegraphics[width=.7\linewidth,clip]{deltad.ps}
\end{minipage}
\vskip-.4cm
\caption{Current errors on polarized parton distributions (left, from
  ref.~[24]
) and expected errors in the CERN neutrino
  factory scenario (right, from ref.~[9]).
\label{nufppdf}}
\end{figure}
\vspace*{-12pt}
The absolute value of each quark distribution, however, is affected by
the  gluon admixture
eq.~(\ref{anomeq}). Therefore, the precision in the determination of
$a_0$ is set by the accuracy in the knowledge of $\Delta g$. The
determination of the latter at a
neutrino factory would improve somewhat thanks to the accurate
knowledge of scaling violations, but it
would be very significantly hampered by the limited kinematic coverage,
in $Q^2$ and especially
at small $x$. On the other hand, by the time a  neutrino factory comes
into operation $\Delta g$ is likely to have been determined at
collider experiments, 
such as RHIC.\cite{polrev} Hence, hadron colliders and the neutrino
factory have complementary roles.
The same applies to the determination of the shape of individual
parton distributions (figure~4). While structure functions would be
measured very accurately, the dominant uncertainty of quark distributions would come from the polarized gluon.

Less inclusive measurements at a neutrino factory could provide a very clean handle on
individual observables: for example, strangeness could be studied
through charm production, just as in the unpolarized case. Thanks to
high event rates, it would be possible to measure even
elusive quantities such as related to polarized fragmentation, or even
generalized parton distributions.\cite{cernufact}

\section{Do we need a neutrino factory?}
Alternative
high--intensity neutrino sources, such as $\beta$ beams, provide
attractive opportunities  for the study of neutrino oscillations. They
share the same advantages as the neutrino factory, but with a somewhat
lower
flux and much lower energy: e.g. at a $\beta$ beam one would expect
$\sim 10^{18}$ $\beta$ decays per year with an
average $\nu$ energy $\sim 200$~MeV. With such a beam only elastic or
quasi-elastic scattering on nucleons can be performed.
However, effective field theory
results\cite{kapman} relate the matrix elements measured in  low-energy  
scattering to polarized partonic observables:
e.g.
$\langle p| {j^\mu_5}^{Z}|p\rangle
=\Delta u-\Delta d-\Delta s=-\frac{1}{3} a_0+a_3+\frac{1}{3}
  a_8$, up to computable  corrections related to higher-dimensional
  operators. This would allow e.g. a direct determination of $a_8$ and
  thus of the total polarized strangeness, if not the separation of
  $\Delta s$ and $\Delta \bar s$.
The full impact of such
  measurements on high-energy nucleon observables is currently under
  investigation.\cite{polbb} 

In sum, a full determination of the polarized flavour structure of the nucleon
will only be possible at a neutrino factory, with collider data providing
complementary information on the polarized glue. Low-energy facilities
such as $\beta$--beams would have a more limited though not negligible impact.

\vspace*{-12pt}
\section*{Acknowledgments}
I thank F.~Bradamante for inviting me at this stimulating meeting.
This work was completed at KITP
Santa Barbara, supported in part by the National Science Foundation
under grant PHY99-0794.

\end{document}